# Contrast mechanisms in pump-probe microscopy of melanin

**David Grass,**[1,*] **Georgia M. Beasley,**[2,3,4] **Martin C. Fischer,**[1,6] **M. Angelica Selim,**[5] **Yue Zhou,**[1] **and Warren S. Warren**[1,3,6,7,8]

[1]*Department of Chemistry, Duke University, Durham, North Carolina, USA*
[2]*Department of Surgery, Duke University, Durham, North Carolina, USA*
[3]*Duke Cancer Institute, Duke University, Durham, North Carolina, USA*
[4] *Department of Medicine, Duke University, Durham, North Carolina, USA*
[5]*Department of Pathology, Duke University, Durham, North Carolina, USA*
[6]*Department of Physics, Duke University, Durham, North Carolina, USA*
[7]*Department of Biomedical Engineering, Duke University, Durham, North Carolina, USA*
[8]*Department of Radiology, Duke University, Durham, North Carolina, USA*
[*]*david.grass@duke.edu*

**Abstract:** Pump-probe microscopy of melanin in tumors has been proposed to improve diagnosis of malignant melanoma, based on the hypothesis that aggressive cancers disaggregate melanin structure. However, measured signals of melanin are complex superpositions of multiple nonlinear processes, which makes interpretation challenging. Polarization control during measurement and data fitting is used to decompose signals of melanin into their underlying molecular mechanisms. We then identify the molecular mechanisms that are most susceptible to melanin disaggregation and derive false-coloring schemes to highlight these processes in biological tissue. We exemplary demonstrate that false-colored images of a small set of melanoma tumors correlate with clinical concern. More generally, our systematic approach of decomposing pump-probe signals can be applied to a multitude of different samples.

## 1. Introduction

In this letter we explore contrast mechanisms provided by pump-probe, transient absorption microscopy to study melanin. Melanin is a natural pigment that is responsible for coloration of human skin and hair as well as protection from UV radiation [1–3]. Pump-probe images of melanin may also provide a biomarker for diagnosis of metastatic melanoma [4, 5]. Melanoma is the cancer of melanocytes, the cells that produce the natural pigment melanin in the epidermal layer of the skin. Thus, melanin is naturally abundant in most melanomas and offers itself as biomarker. Melanoma is the most aggressive skin cancer and the 5th deadliest cancer in the Untied States [6]. Most melanomas can be diagnosed by visual inspection, taking a biopsy, evaluation of histopatholoy and treated by excision. If histopathology identifies a tumor as high risk, sentinel lymph node biopsy is standard evaluation but 85% of melanoma patients who undergo sentinal lymph node biopsy have a negative result. There remains a clinical need to more clearly identify which patients with primary melanomas will develop metastases so that appropriate therapy can be given to reduce recurrence [7, 8]. The development of biomarkers for *metastatic* melanoma is currently an active field of research, but still in its early stage [8–12]. We have already demonstrated that pump-probe microscopy images of primary tumors correlate with metastatic disease [4, 5, 13, 14]. In this letter we present measurement and data evaluation methods for pump-probe microscopy to quantify and visualize the assembly structure of melanin.

Our current hypothesis is that the harsh extra- and intracellular environment of metastatic melanoma alters the melanin assembly structure, from intact, highly assembled structures towards its fundamental monomeric building blocks. To this end we use synthetic melanin (particles) and disaggregated synthetic melanin (subunits) as surrogate endpoints to explore their molecular dynamics. In particular, we focus on measurable differences between melanin and disaggregated melanin and on robust strategies to quantify them. We then apply these separation strategies to natural melanin of a small set of melanoma tumor biopsies and discuss their biological implications.

## 2. Pump-probe microscope and pump-probe characteristics of melanin

Pump-probe microscopy is a combination of laser scanning microscopy and pump-probe spectroscopy [15]. Absorption of a pump pulse in the sample lifts electronic population from the ground state into higher states. The excited states as well as the hole in the ground state evolve with different relaxation times. They are detected with a probe pulse as a function of time delay $t$ between pump and probe. The relaxation times are specific to molecular species and are at the origin of contrast and chemical specificity of pump-probe microscopes. Figure 1a) shows a schematic of the pump-probe microscope. A master oscillator (Coherent Chameleon) and a optical parametric oscillator (Coherent MIRA-OPO) provide pump and probe pulse trains at a repetition rate of 80MHz. The wavelength of pump (probe) pulse train $\lambda_{pump}$ ($\lambda_{probe}$) can be selected from the visible and NIR range. A delay stage is used to control the arrival time $t$ between pump and probe at the sample. Pump and probe pulse trains are superimposed with a beam splitter (BS) and travel co-linearly into a laser scanning microscope. An Olympus 20X objective (MO) is used to focus both lasers onto the sample. Pump-probe signals are measured with a modulation transfer scheme: The pump pulse train is modulated at a frequency of $\Omega_{mod} = 2$MHz with an acousto-optical modulator (AOM). The nonlinear interaction between pump pulse, probe pulse, and sample transfers modulation from the pump pulse train to the probe pulse train, hence modulation transfer. After the interaction with the sample, the pump pulse train is rejected with an optical filter and the probe pulse train is detected with a photodiode. A lock-in amplifier is used to measure differences in probe absorption conditioned on the presence of a pump pulse. Imagine a two-photon absorption event for example: If both, pump and probe are interacting with the sample, one photon of each pulse is absorbed. If the pump is switched off by the AOM, no probe photon is absorbed and a higher probe pulse train intensity is detected. Thus, the two-photon absorption event manifests as loss of one photon in the probe pulse at the modulation frequency $\Omega_{mod}$. A transient absorption curve is acquired by scanning the inter-pulse delay $t$. Our experimental convention displays a loss mechanism, such as two-photon absorption, as a positive signal and a gain mechanism as negative. Pump and probe pulse trains are raster scanned over the sample with scanning mirrors and form pump-probe stack, which is a a three-dimensional data cube with two spatial and one temporal dimension. Note that our microscope allows for change of pump (probe) wavelength $\lambda_{pump}$ ($\lambda_{probe}$), pump (probe) intensity $I_{pump}$ ($I_{probe}$) and the relative polarization angle $\theta$ between pump and probe with a half waveplate ($\lambda/2$).

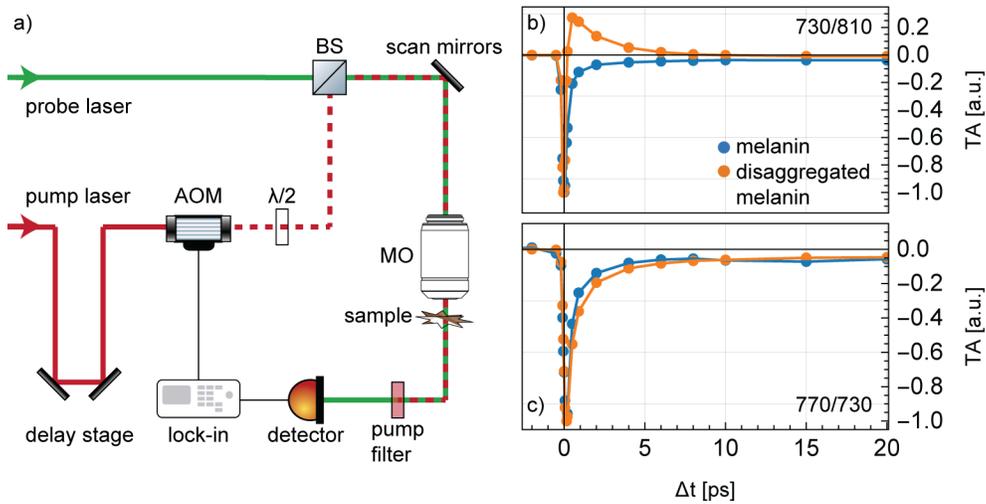

Fig. 1. Pump-probe microscope setup a) and comparison between transient absorption curves of synthetic melanin and disaggregated melanin b), c). For a pump wavelength of $\lambda_{pump} = 730$nm and a probe wavelength of $\lambda_{probe} = 810$nm, disaggregated melanin (orange curve) has an additional positive component b) and for a pump wavelength of $\lambda_{pump} = 770$nm and a probe wavelength of $\lambda_{probe} = 730$nm, intact melanin (blue curve) shows a faster decay c).

In this manuscript we use two wavelength combinations to image melanin: a pump wavelength of $\lambda_{pump}$ = 770nm and a probe wavelength of $\lambda_{probe}$ = 730, abbreviated with 770/730 for the remaining manuscript. The second combination uses a pump wavelength of $\lambda_{pump}$ = 730nm and a probe wavelength of $\lambda_{probe}$ = 810nm, abbreviated with 730/810. The predominant molecular mechanisms for melanin and the two wavelength combinations are ground state bleaching (GSB) and excited state absorption (ESA), indicated by the positive and negative nature of the signals shown in figure 1b) and c). In an ESA event, pump and probe consecutively excite population from the initial state over an intermdiate state to a final state. This loss process appears as a positive transient and decays with the population lifetime of the final excited state. GSB is a competition between pump and probe pulse for population in the ground state and is a transient gain process. It appears as a negative signal. Transient absorption curves of melanin particles and melanin subunits are shown in figure 1. Part b) displays both samples measured with 730/810. The dynamics of melanin are governed by GSB (negative) while an additional ESA (positive) component appears for subunits. Figure 1c) displays measurements with 770/730: Both samples are governed by GSB, but melanin particles decay faster as subunits. These observations agree with previous, qualitative results [4, 5]. In the remaining manuscript we will explore new and more robust strategies to quantify these differences between particles and subunits. First, we developed a fitting model to quantify lifetimes of different molecular mechanisms and second, we harness the dependence of molecular mechanisms on polarization angle $\theta$ between pump and probe. Both approaches are used to derive false-colored images of biological samples at the end of this manuscript.

We selected 730/810 and 770/730 as imaging modalities based on previous work from our group [4, 5], which also hypothesized that pump-probe images contain information about melanin assembly structure and that melanin assembly structure correlates with clinical concern [5]. The previous work used qualitative techniques based on phasor analysis and principle component analysis to classify pump-probe signals. Here, we go one step further and provide techniques to identify the underlying molecular mechanisms that comprise the complex pump-probe signals of melanin. This allows us to unambiguously assign molecular mechanisms to characteristics of the sample under investigation, which was not possible in the previous approaches.

## 3. Fitting

Pump-probe microscopes measure the temporal evolution of ground state holes and excited state populations as a function of inter-pulse delay $t$. A measured pump-probe signal consists of a superposition of multiple nonlinear processes. We model each individual mechanism as an exponential decay $ae^{-t/\tau}$ with a characteristic amplitude $a$ and lifetime $\tau$. The number of excited electrons $n(t)$ after pump absorption is given by the convolution of the pump pulse $I_{pump}(t)$ with the exponential decay of the electronic state $n(t) \propto \int dt' I_{pump}(t - t')ae^{-t'/\tau}$. The population $n(t)$ is measured by the probe pulse and the resulting transient absorption signal $s(t)$ is the

convolution between the excited state population $n(t)$ and the probe pulse $I_{\text{probe}}(t)$

$$s(t, \tau, a) \propto \int_{t'=-\infty}^{t} dt' \, I_{\text{probe}}(t - t') n(t') = \frac{a}{2} e^{-\frac{t}{\tau} + \frac{\beta^2}{4\tau^2}} \left[ 1 + \text{erf}\left(\frac{t}{\beta} - \frac{\beta}{2\tau}\right) \right]. \quad (1)$$

The temporal envelope of pump and probe pulse are modeled as Gaussians with standard deviation $t_{\text{pump}}$ and $t_{\text{probe}}$, respectively, and with the relation $\beta^2 = t_{\text{pump}}^{-2}/2 + t_{\text{probe}}^{-2}/2$. If the lifetime of a molecular process is much shorter than the pulse width of pump and probe pulse, the transient is dominated by the pulse widths of pump and probe. Such a process cannot be temporally resolved and appears as instantaneous.

Typical pump-probe curves, such as shown in figure 1, are a superposition of multiple processes. Exponential functions do not form an orthogonal basis set and cannot unambiguously be separated. There exist limitations on lifetime and amplitude extraction for multi-exponential decays [16, 17] and this situation is aggravated by the bipolar nature of pump-probe signals. We therefore take a careful approach in developing a fit model for pump-probe data to avoid over-fitting.

In order to separate the complex pump-probe signals of melanin we developed fitting models for the data presented in figure 1b) and c). The pump-probe dynamics for melanin subunits acquired with 730/810, see orange data in figure 1b), have at least one ESA and one GSB component as the signal evolves from negative to positive (bipolar signal). The other three curves (negative signals) shown in figure 1 are all dominated by a single or multiple GSB (negative) mechanisms. We use the Akaike information criterion corrected for small sample size (AICc) estimator to determine the optimal number of molecular processes to describe the data [18–20]. AICc measures the quality of fit by using a maximum likelihood estimate and penalizes for number of free parameters. The lower the AICc value, the better the model. We vary the number of molecular processes $s(a, \tau)$ in the fit function and and choose the model that has the lowest AICc.

We begin by developing a fit model for the bipolar pump-probe measurements of melanin subunits acquired with 730/810. The simplest fit function to start with is $g_1 = s_1(a_{\text{GSB1}}, \tau_{\text{GSB1}}) + s_2(a_{\text{ESA1}}, \tau_{\text{ESA1}})$ containing one GSB molecular mechanism $s_1$ for the negative contribution and one positive for the ESA component $s_2$. We test this model on melanin subunits and find a lifetime of $\tau_{\text{GSB1}} \approx 16$ fs for the GSB process. This process is significantly shorter than the pulse width of pump and probe $t_{\text{pump}} \approx t_{\text{probe}} \approx 70 \, fs$ and we cannot experimentally resolve the lifetime of this component, nor distinguish if this decay contains multiple components. It is treated as a single molecular GSB mechanism in the fit model. The next more complicated model we compare $g_1$ with is $g_2 = g_1 + s_3(a_{\text{ESA2}}, \tau_{\text{ESA2}})$ with an additional ESA component. The AICc scores are consistently lower for the simpler model $g_1$. Averaged over five measurements, the probability of the second model $g_2$ to be better as model $g_1$ is $p = \exp((\text{AICc}_1 - \text{AICc}_2)/2) = 0.02$. The result of the AICc analysis matches our intuition: The two fits $g_1$ and $g_2$ and a measurement are shown in figure 2a). The fit functions look almost identical to the eye, but $g_2$ contains two more free fit parameters and is therefore more susceptible to over-fitting. The three remaining measurements shown in figure 1 are qualitatively similar: They have a component that does not visibly decay over 20ps which is described as a molecular component with an infinite lifetime $\tau = \infty$. The most minimalistic model is a combination of the long-lived mechanism and one GSB mechanism to account for the initial decay $f_1 = s_1(a_{\text{GSB1}}, \tau_{\text{GSB1}}) + s_2(a_2, \infty)$. We compare it with the models $f_2 = f_1 + s_3(a_{\text{GSB2}}, \tau_{\text{GSB2}})$ and $f_3 = f_2 + s_4(a_{\text{GSB3}}, \tau_{\text{GSB3}})$ with one or two additional GSB components, respectively. We repeat the same routine as described above for the three remaining cases: 730/810 melanin particles, 770/730 melanin particles and 770/730 melanin subunits. The AICc score is consistently lowest for the bi-exponential model $f_2$. The probability of the mono-exponential model $f_1$ to be better is $p < 0.01$, $p = 0.04$ and $p < 0.01$ for 730/810 melanin particles, 770/730 melanin particles and 770/730 melanin subunits, respectively.

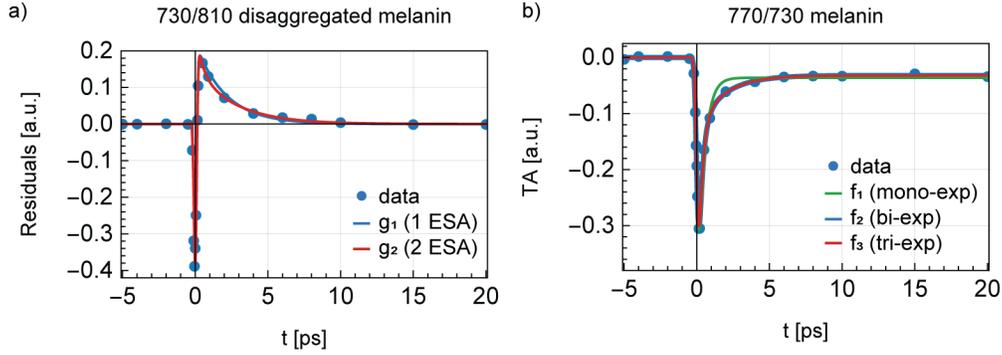

Fig. 2. Fit model development: a) comparison between fit functions $g_1$ and $g_2$ to 730/810 measurements of disaggregated melanin, $g_1$ contains one more ESA process. b) comparison between mono-exponential $f_1$, bi-exponential $f_2$, and tri-exponential $f_3$ fit to 770/730 measurements of melanin. The Akaike information criteria selects $g_1$ and $f_2$ as best choice for fitting.

The probability of the tri-exponential model $f_3$ to be the better model is is $p < 0.01$, $p < 0.01$ and $p < 0.01$ for 730/810 melanin particles, 770/730 melanin particles and for 770/730 melanin subunits, respectively. The AICc analysis agrees again with our intuition. Figure 2b) displays all three fit functions $f_1$, $f_2$ and $f_3$ for melanin particles imaged with 770/730. The first fit function (green curve) visibly deviates from the data between 0 and 5ps. The fit functions $f_2$ and $f_3$ overlap almost perfectly. Therefore, the AICc score is higher for $f_3$ as it contains two more free fit parameters than $f_2$.

| sample | $\lambda$ [nm] | $\tau_{GSB1}$ [ps] | $\tau_{GSB2}$ [ps] | $\tau_{ESA1}$ [ps] | $\tau_{ESA2}$ [ps] |
|---|---|---|---|---|---|
| particle | 730-810 | $(16 \pm 1) \times 10^{-3}$ | $0.67 \pm 0.13$ | NA | NA |
| subunit | 730-810 | $(16 \pm 1) \times 10^{-3}$ | NA | $2.11 \pm 0.16$ | $0.34 \pm 0.12^*$ |
| particle | 770-730 | $0.13 \pm 0.01$ | $1.21 \pm 0.21$ | NA | NA |
| subunit | 770-730 | $0.23 \pm 0.02$ | $2.13 \pm 0.21$ | NA | NA |

Table 1. Molecular processes and their lifetimes for synthetic melanin and subunits. Note that processes that have a much longer lifetime compared to the acquisition time ($\tau \gg 20$ps) are not displayed in this table.* only detectable with polarization measurement, see next chapter.

In summary, the AICc analysis identifies $g_1$ (GSB + ESA) as best fit model for subunits imaged with 730/810 and $f_2$ (bi-exponential) as best fit model for all other cases. The results from fitting are summarized in table 1 and reveal a good candidate to distinguish melanin from disintegrated melanin: The lifetimes of the GSB processes for a pump wavelength of $\lambda_{pump} = 770$ nm and a probe wavelength of $\lambda_{probe} = 730$ nm. The long-lived offset signals are not included in this summary. All measurements were performed with a relative polarization angle of $\theta = 0$ between pump and probe pulse. Note that there exists an additional ESA component for melanin subunits imaged with 730/810 that is only observable for relative polarization angles approaching $\theta = \pi/2$. Its lifetime is denoted with a * symbol in table 1 and the process is described in more detail in the next section.

## 4. Polarization dependence

We recently reported polarization modulation pump-probe microscopy and demonstrated that polarization modulation can be used to experimentally separate two molecular species [21]. Here we use a similar approach, based on a simple dipole transition model, to utilize polarization dependence for identification of molecular processes in melanin. The amplitude of a molecular process as a function of relative polarization angle between pump and probe pulse $\theta$ is modeled as

$$a(\theta) \propto b + c \cos\left[2(\theta - \alpha_1)\right] \quad (2)$$

with $b, c$ and $\alpha$ constants that are specific to the molecular mechanism. If a superposition of multiple mechanisms $s(\theta) = \sum_j s(t, \tau_j, a_j(\theta))$ with distinct polarization dependencies is measured with $n$ different polarization angles $\theta_i$, each individual mechanism can be reconstructed by solving the resulting system of $n$ linear equations $s(\theta_i) = \sum_j s(t, \tau_j, a_j(\theta))\big|_{i=1,\dots,n}$. Only a-priori knowledge of the polarization dependencies is required. Here, the a-priori knowledge is obtained by measuring homogeneous samples of melanin particles and melanin subunits for seven equidistant polarization angles between $\theta = 0$ and $\theta = \pi/2$. Each transient absorption curve is fitted to the model described in the previous section and the results are summarized in figure 3. The data points represent the amplitude free fit parameters, the shaded areas represent fit uncertainty and the solid curves are fits to equation 2.

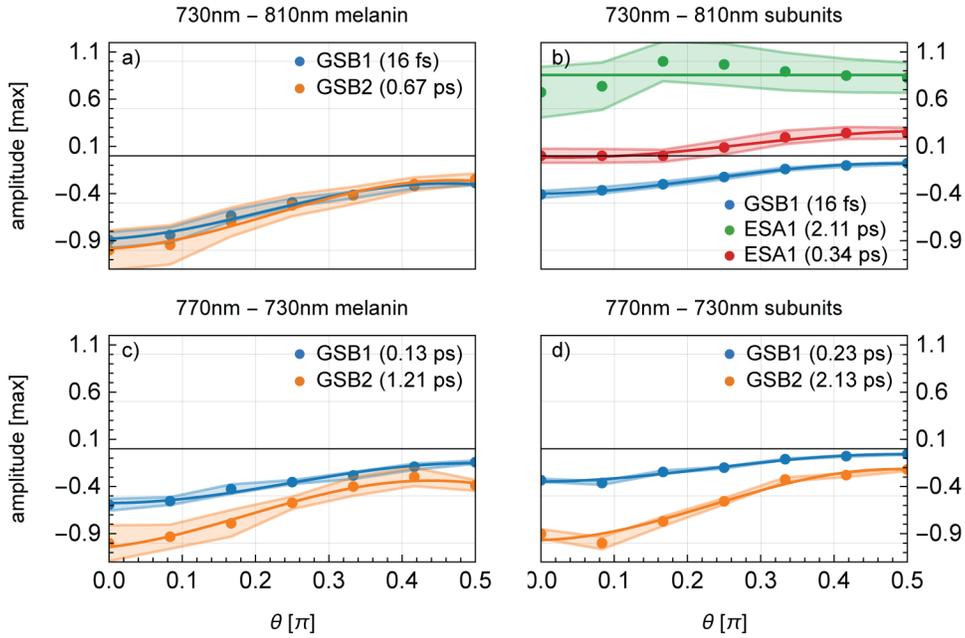

Fig. 3. Individual molecular mechanisms of melanin and disaggregated melanin as a function of relative polarization $\theta$ between pump and probe pulse. The number in brackets represents the lifetime of each individual mechanism. Solid points represent amplitudes from fitting, shaded areas represent fitting uncertainty and solid curves represent fits to polarization dependence equation 2.

For melanin subunits imaged with 730/810, see figure 3b) we found an additional ESA component (red curve) that is not observable for parallel polarization and becomes strongest for a relative polarization angle of $\pi/2$. We used again the AICc criteria to determine that a single additional ESA component is sufficient to describe the measurement. In the same measurement, ESA1 only shows a weak polarization dependence, see green curve in figure 3b). The probability of a polarization dependent model to be better than a polarization independent model ($c = 0$ in equation 2) is $p = 0.001$. Based on this AICc evaluation we choose to model the ESA1 process as polarization independent. The results of the polarization dependence for 730/810 are summarized in table 2. Note that the coefficients are normalized to the value of $b$ and a standard Gaussian error propagation was used to determine the uncertainties. Additionally, an AICc evaluation of the polarization dependence for all molecular processes, including 770/730, can be found in the supplementary information A .

| process | sample | $\lambda$ [nm] | $b[b]$ | $c[b]$ | $\alpha[\pi]$ |
|---|---|---|---|---|---|
| GSB1 | particle | 730-810 | $1.00 \pm 0.04$ | $0.50 \pm 0.05$ | $-0.04 \pm 0.02$ |
| GSB2 | particle | 730-810 | $1.00 \pm 0.04$ | $0.59 \pm 0.06$ | $-0.03 \pm 0.02$ |
| GSB1 | subunit | 730-810 | $1.00 \pm 0.02$ | $0.66 \pm 0.03$ | $-0.02 \pm 0.01$ |
| ESA1 | subunit | 730-810 | 1 | 0 | NA |
| ESA2 | subunit | 730-810 | $1.00 \pm 0.14$ | $-1.13 \pm 0.26$ | $0.03 \pm 0.03$ |

Table 2. Polarization dependence of individual molecular mechanisms for melanin and disaggregated melanin imaged with 730/810.

The GSB1 component for particles and subunits, measured with 730/810, has the same lifetimes, see table 1, and a similar polarization dependence. Subunits are derived from melanin particles by exposure to an alkaline solution and contain the same building blocks as particles. Thus, pump-probe signals originating from the fundamental buildings blocks would appear in both samples and it is very likely that GSB1 is the same molecular process in both samples. GSB1 and GSB2 also have a very similar polarization dependence and for the remainder of the manuscript we approximate their polarization dependence to be the same. We will later use polarization properties to separate the ESA1 molecular process from the GSB processes. The polarization dependencies for 770/730 do not reveal obvious candidates to separate particles from subunits and the results can be found in the supplementary information A.

## 5. False-coloring schemes

Our goals are development of imaging and data evaluation strategies to extract biological relevant contrast of melanin samples. Ultimately, we want to derive a diagnostic biomarker for metastatic melanoma based on pump-probe images of primary tumor samples. In a first step, we want to derive false colored images based on melanin assembly state. In the previous two chapters we used fitting and polarization to characterize molecular mechanisms of synthetic melanin and its disaggregated version.

The synthetic melanin particles were synthesized by spontaneous oxidation of dopamine [22] and a part was disaggregated with deoxygenated alkaline solution [23] into subunits. In biological systems, such as humans, melanin degradation is a continuous process that can have many origins such as natural degradation due to aging, sun exposure, digestion by melanophages in addition to our hypothesis that aggressive melanomas alter melanin structure. Therefore we expect to find a spectrum ranging from intact melanin over partially disintegrated to fully disintegrated in natural systems.

The most obvious difference between melanin particles and subunits is the presence (or absence of) the ESA1 component in images of melanin subunits (particles) acquired with 730/810. We use polarization dependent measurements to quantify GSB and ESA processes individually and derive a contrast based on their ratio. This has the advantage that the result will be independent of the overall amount of melanin present in the sample. We assume the polarization dependence of GSB1 (for both particles and subunits) and GSB2 to be the same, see free fit parameters in table 2, with $b = 1, c = 0.5$ and $\alpha = 0$, which reflects the polarization dependence of GSB in a simple dipole transition model [21]:

$$\begin{aligned} s_{\text{meas}}(t, \theta) &= \frac{1}{2}\left\{2 + \cos\left[2(\theta - \alpha)\right]\right\}\left[s(t, \tau_{\text{GSB1}}) + s(t, \tau_{\text{GSB2}})\right] + s(t, \tau_{\text{ESA1}}) + s(t, \tau_{\text{ESA2}}, \theta) \\ &= \frac{1}{2}\left\{2 + \cos\left[2(\theta - \alpha)\right]\right\}s(t, \tau_{\text{GSB}}) + s(t, \tau_{\text{ESA1}}) + s(t, \tau_{\text{ESA2}}, \theta) \end{aligned} \quad (3)$$

with $s(t, \tau_{\text{GSB}})$ the combination of all GSB processes. This leaves three molecular mechanisms and, in principle, three measurements with different relative polarization angle would be sufficient to unambiguously separate them. In order to minimize laser exposure to the sample and reduce experimental complexity, we limit ourselves to two measurements with parallel and perpendicular polarization and assume that the amplitude of the ESA2 process is negligible with respect to ESA1 and the GSB processes. Using these assumptions and the polarization independence of ESA1 allows reconstruction of the GSB processes and the ESA1 process as

$$\begin{aligned} s_{\text{GSB}}(t) &\approx \frac{3}{2} s_{\text{meas}}(t, 0) - \frac{3}{2} s_{\text{meas}}(t, \frac{\pi}{2}) \\ s_{\text{ESA1}}(t) &\approx -\frac{1}{2} s_{\text{meas}}(t, 0) + \frac{3}{2} s_{\text{meas}}(t, \frac{\pi}{2}). \end{aligned} \quad (4)$$

In a last step we integrate the absolute value of the GSB and ESA1 signals over the acquisition time and compute their ratio. A higher numerical value suggests more disaggregated melanin. As this metric is based on a ratio, it contains an intrinsic normalization that allows comparison between different samples which may differ in the amount of melanin.

As a second metric to distinguish melanin particles from disaggregated melanin we select the lifetime $\tau_{\text{GSB2}}$ of 770/730 images. To derive the false-colored images we fit each pixel of a pump-probe stack to the model described earlier and use the lifetime $\tau_{\text{GSB2}}$ for false-coloring. We choose $\tau_{\text{GSB2}}$ over $\tau_{\text{GSB1}}$ as it is much longer than the pulse width of pump and probe pulse and hence can be better temporally resolved. Longer lifetimes are expected for disaggregated melanin and note that this metric is also concentration independent.

## 6. False-colored pump-probe images of melanoma

We selected a small set of representative melanoma biopsies with clear clinical diagnosis: A patient sample with melanin from a local melanoma that did not evolve into metastatic cancer (follow up >10 years), melanin in macrophages, melanin in a lymph node from metastases and melanin in a primary tumor that has evolved into metastatic cancer. These samples were selected under the premises that they should contain the two endpoints of melanin assembly state: intact melanin and disaggregated melanin. The samples used for imaging are 2 adjacent, $5\mu$m thick cuts from tumor tissue fixed with formalin and embedded in paraffin. The first slice is stained with hematoxylin and eosin (HE) that is commonly used for melanoma diagnosis by pathology. The HE slice serves as reference map to compare to pump-probe images. The second slice is deparaffinized by baking it in an oven to melt paraffin and consecutive washing with xylene and graded alcohols. Deparaffinization reduces scattering of probe photons in the sample and allows for a higher dynamic range with the lock-in detection. In a typical experiment, pathology marks melanoma on the HE slide. Then, areas that contain melanin within the tumor on the unstained slide are imaged with the pump-probe microscope. For 770/730, the polarization angle between pump and probe is set to $\theta = 0$ and one pump-probe stack of the region of interest is imaged. For 730/810 imaging, two pump-probe stacks of the same region of interest are acquired: first with a relative polarization angle of $\theta = 0$ and a second run with $\theta = \pi/2$.

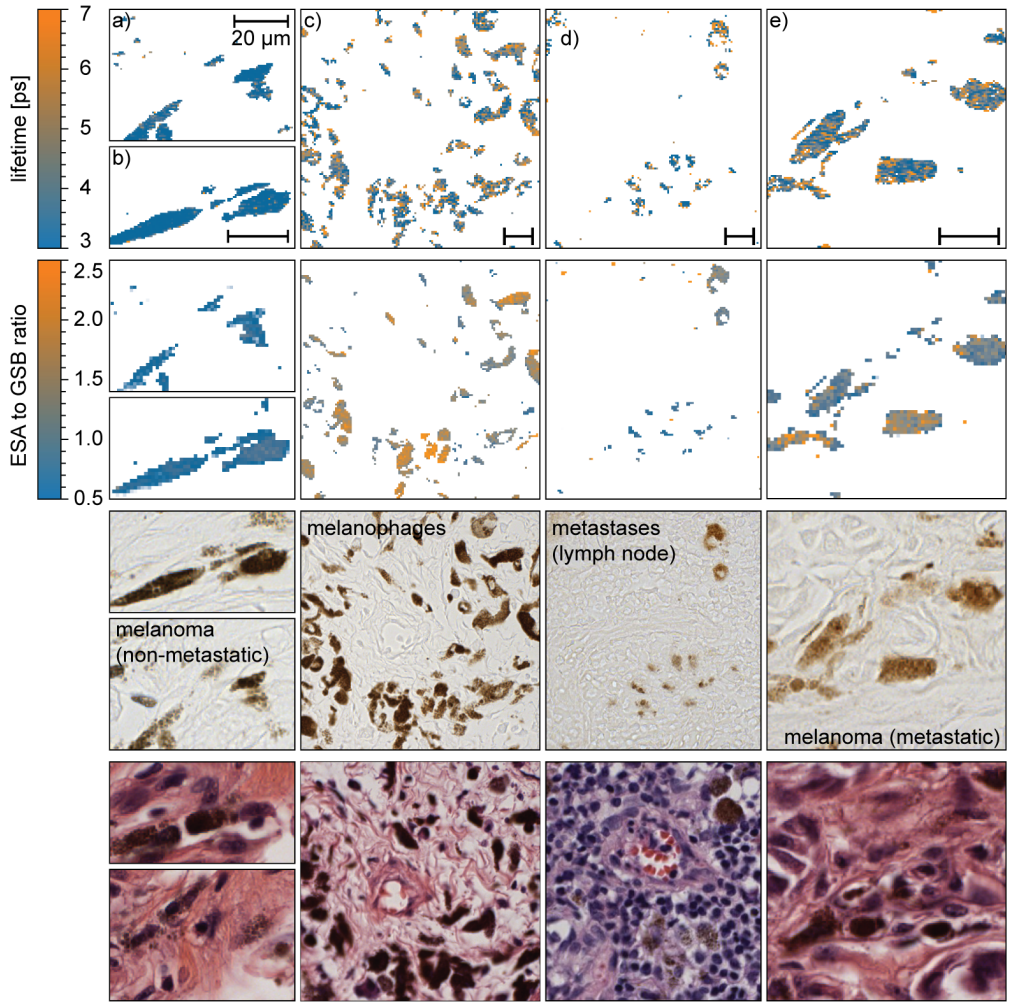

Fig. 4. Lifetime and ESA to GSB ratio images of melanoma samples derived from 770/730 and 730/810. The images show melanin from a a)-b) non-metastatic, local tumor, c) melanophages, d) lymph node and e) primary tumor that evolved into metastatic cancer. The first two rows display the false-colored, pump-probe images, the third row shows brightfield images of the unstained slide and the fourth row shows brightfield images of the adjacent HE slide. Scale bar: 20$\mu$m

Figure 4 shows the microscopic images of the four tumors. The top row displays false-colored, pump-probe lifetime images, the second row displays false-colored, pump-probe images that highlight the ESA to GSB ratio, the third row displays brightfield images of the unstained tumor and the bottom row shows brightfield images of the same area from the adjacent HE slide. The melanin (brown-ish color) in parts a)-b) is from two regions within the same thick primary tumor (Breslow thickness of $d = 6$mm) located deep inside the dermis. Both false-coloring schemes suggest that the melanin of this tumor is predominantly intact. Image c) displays melanin found in a non-metastatic tumor, but most of the melanin is located inside melanophages. Melanophages are macrophages that ingest melanin, and macrophages are cells of the immune system that digest pathogens to protect their host. Thus, melanin found in and around melanophages is very likely disaggregated, which is suggested by the orange color of the lifetime and the ESA to GSB ratio image. Image d) displays melanin in a lymph node. This melanin is produced from melanocytes inside the lymph node, which are metastasis that mirgrated from a primary melanoma through the lymphatic system. Such tumors are the most dangerous and we would expect a strong signature in melanin. Both false-colored pictures indicate that the melanin is disaggregated, but interestingly, there are also areas that indicate intact melanin. Part e) displays melanin from a primary tumor that developed metastatic disease. Both false-coloring schemes indicate disaggregated melanin. These five images show various degrees of melanin deaggregation that match expectations from clinical diagnosis. A non-metastatic tumor is less aggressive than a metastatic tumor. Thus, we would expect more disaggregated melanin in the lymph node c) and the metastatic tumor d) compared to a local, non-metastatic tumor shown in part a). Similarly, pump-probe images show that melanin found in melanophages is disaggregated.

## 7. Future work and conclusions

Pump-probe microscopy generates decays in melanin that show substantial differences (both in lifetimes and in polarization dependence) with disaggregation, which has previously been suggested as an apparent hallmark of metastatic melanoma [4, 5]. The dynamics remain rich and complex, reflecting contributions from multiple physical mechanisms. Polarization pump-probe microscopy in combination with fitting gives images of biopsy specimens with significant contrast, that at least in the few cases examined here correlate with clinical concern. We used synthetic melanin and its disaggregated form as surrogate for natural melanin and project the differences onto biological systems. The conclusions that can be drawn from this approach are limited by the accuracy of the synthetic model. The synthetic melanin used in this manuscript is a robust model, however it does not capture all aspects of melanin in organisms. For example, melanin is also a chelation agent for metals and we know from previous studies that metal content in melanin affects pump-probe signals [24]. Thus, we need to develop and use more complex melanin models to closer simulate in-vivo conditions, in order to better understand melanin and its relation to melanoma. Nonetheless, this work makes it clear that the pump-probe decay can be separated into its underlying molecular mechanisms, which for melanin samples do differ significantly in degraded samples. We also show that melanin degradation produces image characteristics also seen in the comparison between biopsies from metastatic and non-metastatic lesions. This is a highly plausible result as of the harsh extracellular and intracellular environment from metabolically aggressive tumors. However, this is not a clinical study; we solely want to demonstrate that the developed data acquisition and evaluation strategies work on realistic examples and are ready to be used for a clinical study.

**Funding.** funded by the Deutsche Forschungsgemeinschaft (DFG, German Research Foundation) – 450198538; Chan Zuckerberg Initiative (2021-242921); National Science Foundation (CHE-2108623); Georgia M Beasley supported by NIH K08 CA237726-01A1;

**Acknowledgement.** We thank Shannon Eriksson, Jacob Lindale, Simone Degan, Jennifer Zhang, Suephy Chen, Chuan-Yuan Li, Russell Hall and Meenal Kheterpal for stimulating discussions.

## A. Supplementary Information: Polarization dependence

Table 3 shows the fit results of the polarization characterization of melanin particles and melanin subunits to equation 2.

The AICc is used for every molecular mechanism to distinguish if they have a polarization dependence or if they are better described by a polarization independent model. The results are summarized in table 4. $AICc_1$ represents the AICc score for a polarization dependent model and $AICc_2$ for a polarization dependent model. $p$ represents the probability of the model with the higher AICc score to describe the data better. The only mechanism that is not polarization dependent is ESA1 for the case of 730/810 imaging. All other molecular mechanisms are better described by a polarization dependent model.

| process | sample | $\lambda$ [nm] | $b$ [$b$] | $c$ [$b$] | $\alpha$ [$\pi$] |
|---|---|---|---|---|---|
| GSB1 | particle | 730-810 | 1.00 ± 0.04 | 0.50 ± 0.05 | −0.04 ± 0.02 |
| GSB2 | particle | 730-810 | 1.00 ± 0.04 | 0.59 ± 0.06 | −0.03 ± 0.02 |
| GSB1 | subunit | 730-810 | 1.00 ± 0.02 | 0.66 ± 0.03 | −0.02 ± 0.01 |
| ESA1 | subunit | 730-810 | 1 | 0 | NA |
| ESA2 | subunit | 730-810 | 1.00 ± 0.14 | −1.13 ± 0.26 | 0.03 ± 0.03 |
| GSB1 | particle | 770-730 | 1.00 ± 0.04 | 0.58 ± 0.07 | 0.00 ± 0.02 |
| GSB2 | particle | 770-730 | 1.00 ± 0.03 | 0.52 ± 0.05 | −0.06 ± 0.01 |
| GSB1 | subunit | 770-730 | 1.00 ± 0.06 | 0.72 ± 0.10 | −0.01 ± 0.02 |
| GSB2 | subunit | 770-730 | 1.00 ± 0.06 | 0.63 ± 0.10 | −0.02 ± 0.02 |

Table 3. Polarization dependence of molecular components of melanin and disaggregated melanin

| process | sample | $\lambda$ [nm] | $AICc_1$ | $AICc_2$ | $p$ |
|---|---|---|---|---|---|
| GSB1 | particle | 730-810 | −18.833 | −10.789 | 0.018 |
| GSB2 | particle | 730-810 | −15.973 | −7.651 | 0.016 |
| GSB1 | subunit | 730-810 | −38.724 | −18.348 | < 0.001 |
| ESA1 | subunit | 730-810 | −5.690 | −19.173 | 0.001 |
| ESA2 | subunit | 730-810 | −18.348 | −15.929 | 0.298 |
| GSB1 | particle | 770-730 | −30.554 | −25.818 | 0.094 |
| GSB2 | particle | 770-730 | −22.681 | −14.339 | 0.015 |
| GSB1 | subunit | 770-730 | −6.991 | −1.113 | 0.053 |
| GSB2 | subunit | 770-730 | 5.572 | 8.847 | 0.194 |

Table 4. AICc criteria for polarization dependence of all molecular processes. $AICc_1$ ($AICc_2$) represents the AICc score for a polarization dependent (polarization independent) model and the $p$ represents the probability of the model with the higher AICc score to describe the data better.